# Concussion and head acceleration exposure in elite rugby union and American football: Interpreting indicative trends

Dr Gregory Tierney[1]

[1] Nanotechnology and Integrated Bioengineering Centre (NIBEC), School of Engineering, Ulster University, Belfast, United Kingdom

Corresponding author: Dr Gregory Tierney, Ulster University, Belfast, United Kingdom.

Email: g.tierney@ulster.ac.uk

**Abstract**

Elite-level American football and rugby union are two high-contact sports with growing clinical and legal concerns over player safety. A comparison of current indicative trends in concussion and head acceleration events (HAEs) in elite-level American football and rugby union was undertaken. Rugby union players have a greater number of professional playing years and matches available in a season than their American football counterparts. Rugby union players have a greater number of concussions reported per match and a higher proportion of concussions occurring during training sessions, based on National Football League (NFL) and Rugby Football Union (RFU) injury reports. Preliminary findings indicate that rugby union forwards experience a higher incidence of HAEs per player match over lower and higher magnitude thresholds, than American football defensive players. Taken collectively, current evidence indicates a trend toward greater reported exposure in elite-level rugby union across several key metrics. These patterns highlight the need for independent, transparent, and scientifically rigorous research to better understand concussion and HAE biomechanics and to support the development of effective mitigation strategies in contact sport.

## 1. Introduction

The physical and high impact nature of certain contact sports has made concussion and repetitive head acceleration event (HAE) exposure a concern.[1] The concerns stem from the potential medium and long-term consequences associated with concussion and repetitive HAE exposure during an athlete's playing career.[1] Sirisena et al. [2] believe the continued interest in this area was potentially fuelled by American football and the National Football League (NFL) players' lawsuit where a group of retired NFL players won a large out-of-court settlement from the NFL, stating they were not warned about the risks associated with concussion and that cases were hidden from them. Rugby union governing bodies are currently involved in a similar lawsuit.[3] The aim of the current article is to conduct a comparison of current indicative trends in concussion and head acceleration events (HAEs) in elite-level American football and rugby union based on governing body reports and available scientific literature. The comparison is limited to the men's games as a direct women's equivalent of elite-level American football data does not exist.[4,5]

## 2. Contextual overview of concussion and HAE metrics

### *2.1. Professional playing career, match format and schedule*

The average professional playing career for rugby union appears to be seven years based on the Irish rugby union players association.[6] For American football, the average professional playing career appears to be 3-6 years.[7-9] The author is unaware of elite-level academy and college playing career length for rugby union and American football players, respectively.

Rugby union matches are typically 80 minutes (excluding 'extra time'),[10] whilst American football matches are typically 60 minutes (excluding 'overtime').[11] In rugby union, each team has 15 players on the field at any given time, made up of forwards and backs.[12] In American football, each team has 11 players from either their 'defense', 'offense' or 'special teams' cohort depending on the match scenario.[13] Accordingly, a minimum of 30 and 44 players are required for a game of rugby union and American football (excluding unique special teams players such as kickers, punters and long-snappers).

Rugby union turned professional in 1995.[10] The first English Premiership Rugby season (1987/88) began with 12 clubs playing each other just once.[14] Since the 1993/94 season, teams play each other twice during the regular season with post-season knockout matches introduced in the 2000/01 season for the top performing teams in the league.[14] There has also been an annual knock-out style cup competition during this time period,[15] and the possibility to play in a European competition since 1995.[16] The inaugural Rugby World Cup took place in 1987 and occurs every four years involving international teams.[17] The Five Nations competition (Six Nations since 2000) was also played annually during this time period. Summer and Autumn international test series have also been common during non-Rugby World Cup years.[18] During Rugby World Cup years, the Autumn international test series is replaced with the Rugby World Cup competition.[19]

In the NFL, a more consistent schedule has existed with a 17 match schedule during the regular season since the 2021/22 season (16 regular season matches pre-2021/22 season) with up to 4 matches possible in the knock-out style post-season.[20] Pre-season schedule typically involves 3 matches per team since the 2021/22 season (4 pre-season matches pre-2020/21 season).[20] However, two teams will play an additional 'Hall of Fame' match each year in the preseason.[20]

Based on Table 1, there appears to be a greater number of matches available to a Premiership rugby union player than a NFL American football player.[18-26] However, Quarrie et al.[27] found that the median number of matches for a Premiership rugby union player in 2014 was 16, with 40% of players appearing in 20 matches or more, 5% appearing in 32 matches or more and 1% appearing in 36 matches or more (Table 2). Table 2 includes non-Premiership matches e.g., international fixtures and cup competitions. The author is unaware of a similar study in the NFL on match appearances for comparison.[27] Match load limits were implemented in Premiership rugby during the 2019/20 season, restricting players from participating in more than 35 matches per season (if playing over 20 minutes) or exceeding 30 full-match equivalents, including non-Premiership matches.[28] In the 2024/25 season, the maximum number of match involvements was set to 30 per season as involvement in 31 or more matches in a season was linked to a significantly higher injury in the subsequent season.[28]

**Table 1.** Example of the available match numbers in professional rugby union in England and a American football in a season.[18-26]

| | Match Numbers | |
|---|---|---|
| | **American Football** | **Rugby Union** |
| **Club** | | |
| Domestic League | National Football League | Premiership Rugby |
| Pre-Season | 5⁺/4⁺⁺ | 3 |
| Regular Season | 16⁺/17⁺⁺ | 22 |
| Post-Season | 4 | 2 |
| Domestic Cup | N/A | Premiership Rugby Cup<br>6 |
| Continent-based Competition | N/A | European Champions Cup<br>9 |
| **International** | | |
| Continent-based Competition | N/A | Six Nations<br>5 |
| World Competition | N/A | Rugby World Cup<br>7* |
| Test Series 1 | N/A | Summer Series<br>4*/3** |
| Test Series 2 | N/A | Autumn Test Series<br>4** |
| Trophy Match | N/A | Quilter Cup<br>1 |
| **Total** | 26 | 59*/55** |

⁺ Applies to NFL up to 2019/20 season.
⁺⁺ Applies to NFL from 2021/22 season.
* Applies to Rugby World Cup years only. 2019 used as reference year.
**Applies to Non-Rugby World Cup years only. 2018 used as reference year (i.e., pre-match limit introduction).

**Table 2.** Percentile breakdown of match appearances by Premiership rugby union players in 2014.[27]

| Percentile | Match Appearances |
|---|---|
| 50 | 16 |
| 60 | 20 |
| 70 | 24 |
| 80 | 27 |
| 90 | 29 |
| 95 | 32 |
| 99 | 36 |

### *2.2. Concussion incidence*

Concussion incidence in rugby union tends to be reported as the number of concussion injuries per 1000 player-hours.[29,30] 1000 player hours equals 25 rugby union matches, excluding 'extra time'.[29] Concussion incidence in American football tends to be reported as Athletic Exposures (AE) or on a per match basis.[30,31] The AE metric provides an overall risk assessment per session of athlete participation.[31] A systematic review of concussion incidence in 'high level' sport, including soccer, rugby union, ice hockey and American Football, found the highest concussion incidences in rugby union match play.[30]

The Rugby Football Union (RFU) produced publicly available injury reports for the seasons 2015/16 to 2022/23.[32] The concussion section of the reports are based on concussion injuries sustained during regular season Premiership, European competition and domestic cup (Anglo-Welsh Cup/Premiership Rugby Cup) matches. The Anglo-Welsh Cup did not take place in 2015/16 season and the cup was replaced by the Premiership Rugby Cup in 2018 (Table 1). Since 2011, concussion is consistently the most reported match injury in the English Premiership.[32,33] In 2022/23, concussion was the most commonly reported match injury, accounting for 24% of all match injuries, a similar trend and value to the 2016-2022 period mean of 23%.[32] Injury reports have also been conducted for previous Rugby World Cup competitions.[34,35] The match concussion incidence was marginally lower in the Rugby World Cup 2019 than the Rugby World Cup 2015 (12.2 v 12.5 injuries per 1000 player hours, respectively).[34,35]

Nathanson et al.[31] found a concussion incidence rate of 0.61 per match in NFL matches over the 2012-2013 and 2013-2014 seasons, excluding preseason matches. Bedard & Lawrence [36] collected injury report data for all NFL injuries from 5 seasons (2012/13-2016/17) and found a concussion incidence rate of 0.61 per match. However, the study included injuries sustained in practice as match injuries and a concussion was defined as any injury reported as a concussion or head injury. Therefore, match concussion incidence may have been overestimated.[36] The NFL releases annual injury data compiled and analysed by an independent third-party.[20] Based on NFL and RFU injury reports, the number of concussion injuries per match appears slightly higher in rugby union than American Football (Table 3). Additionally, a higher percentage of total concussions appear to occur in training practice in rugby union based on NFL and RFU injury reports (Table 4).

**Table 3.** Concussion injuries per match in American Football (regular season matches) and rugby union (Premiership, European competition and domestic cup matches) based on NFL and RFU injury reports.[20,32]

| Season | Rugby Union | American Football |
|---|---|---|
| 2015/16 | 0.63 | 0.71 |
| 2016/17 | 0.84 | 0.65 |
| 2017/18 | 0.72 | 0.70 |
| 2018/19 | 0.82 | 0.50 |
| 2019/20 | 0.79 | 0.53 |
| 2020/21 | 0.89 | 0.50 |
| 2021/22 | 0.73 | 0.46 |
| 2022/23 | 0.74 | 0.55 |
| 2023/24 | - | 0.56 |
| Mean | 0.77 | 0.57 |

**Table 4.** Percentage of total concussions per season from training practice in American football and rugby union based on NFL and RFU injury reports.[20,32]

| Season | Rugby Union | American Football |
|---|---|---|
| 2015/16 | 13.7% | 4.7% |
| 2016/17 | 11.1% | 3.5% |
| 2017/18 | 18.6% | 6.3% |
| 2018/19 | 18.6% | 5.9% |
| 2019/20 | 17.6% | 6.2% |
| 2020/21 | 11.5% | 9.2% |
| 2021/22 | 21.4% | 6.7% |
| 2022/23 | 21.5% | 7.5% |
| 2023/24 | - | 5.6% |
| Mean | 16.8% | 6.2% |

### *2.3. Concussion biomechanics*

The mean peak linear acceleration (PLA) and peak angular acceleration (PAA) in studies reporting specifically on clinically diagnosed concussions in male American football and rugby union has been previously reported.[1] The methodological approaches included instrumented helmets (e.g., Head Impact Telemetry Systems), instrumented mouthguards (iMG; e.g., X2 Biosystems & Prevent Biometric), multibody modelling and Anthropomorphic Test Dummy reconstructions, including reanalysed impacts from earlier studies. Though the head kinematics of concussion appeared similar for both sports, the limited rugby union dataset, wide range and variation of kinematics reported and lack of validity of certain biomechanical approaches undertaken illustrated that a robust comparison was not achievable.[1,37,38]

### *2.4. Video review of contact event incidence*

Quarrie et al.[12] coded the actions of 763 international rugby union players from video recordings of 90 international matches from 2004 to 2010. Table 5 provides details of the contact events of a player per match from tackles, rucks and mauls by positional group. Over the course of a match, Quarrie et al.[12] found that forwards were much more heavily involved in contact events than backs. However, backs are believed to be involved in higher speed collisions than forwards due to the nature of their roles.[39] Smart et al.[40] found from five matches that the mean (± SD) total impacts (sum of tackles made, hit-ups, first three on attack and first three on defence) per player match for forwards was 43.6 (±18.3) and for backs was 13.5 (±7.4). First three players on attack was defined as “being one of the first three players at the breakdown while their team is attacking” and first three players on defence was defined as “being one of the first three players at the breakdown while their team is defending”.[40] The author is unaware of a similar study in American Football utilising match data-feeds or video analysis of contact event incidence for comparison. Attempts have been made to quantify collision incidence in American football using micro-sensor technology embedded in GPS devices worn by players.[41,42] However, the impact zones used in the studies for classifying the impact severity are based on manufacturer recommendations which have never been validated. Additionally, the majority of recordings are believed to be non-contact events as video verification was never used to verify the impacts and thus, it cannot be determined if these are true positive impacts being recorded.[41,42] Lastly, the devices are not rigidly connected to the body and a validation study is needed to assess the accuracy of the acceleration readings from the micro-sensor technology. Consequently, this data will not be used for any comparison.

**Table 5.** Contact events of a player per match from tackles, rucks and mauls by positional group.[12]

| | | **Positional group** | | | | | | | | | |
|---|---|---|---|---|---|---|---|---|---|---|---|
| **Activity** | **Role** | **Prop** | **Hooker** | **Lock** | **Flankers** | **Number 8** | **Scrum-half** | **Fly-half** | **Midfield back** | **Wing** | **Fullback** |
| Scrums | Starter | 21 ± 8 | 20 ± 8.3 | 22 ± 8.1 | 23 ± 7.8 | 23 ± 8.5 | | | | | |
| | Sub | 7.3 ± 4.5 | 6.3 ± 4.6 | 6.8 ± 4.4 | 6 ± 3.9 | 6.6 ± 4.4 | | | | | |
| | Overall | 25 ± 7.8 | 25 ± 7.6 | 25 ± 7.9 | 25 ± 7.9 | 25 ± 7.5 | | | | | |
| Rucks attended - team in possession | Starter | 24 ± 10 | 18 ± 9.3 | 23 ± 11 | 22 ± 8.9 | 20 ± 8.6 | 2.6 ± 2 | 5.4 ± 4.1 | 9.7 ± 5.5 | 7 ± 3.6 | 7.2 ± 4.5 |
| | Sub | 8.1 ± 5.7 | 6.5 ± 5.3 | 8.5 ± 5.9 | 7.2 ± 4.8 | 6.1 ± 4.3 | 1.3 ± 1.4 | 1.9 ± 2.1 | 3.2 ± 3 | 2 ± 1.9 | 2.8 ± 2.6 |
| | Overall | 28 ± 6.7 | 24 ± 7.3 | 28 ± 7.6 | 26 ± 7.2 | 23 ± 6.5 | 3.8 ± 2.6 | 6.4 ± 3.9 | 11 ± 5.3 | 7.7 ± 3.2 | 7.9 ± 4.4 |
| Rucks attended - team not in possession | Starter | 5.9 ± 3.5 | 5.7 ± 3.6 | 6.3 ± 4.2 | 8.7 ± 5 | 7.5 ± 3.9 | 1.8 ± 2 | 2.5 ± 2.3 | 3.4 ± 2.5 | 2.2 ± 1.9 | 1.7 ± 1.7 |
| | Sub | 2.3 ± 2 | 2.1 ± 2 | 2.7 ± 2.4 | 2.8 ± 2.4 | 3.3 ± 2.4 | 0.92 ± 1.3 | 0.96 ± 1.2 | 1.1 ± 1.2 | 0.45 ± 0.81 | 0.68 ± 1 |
| | Overall | 6.8 ± 3.5 | 7.5 ± 3.7 | 7.4 ± 4.2 | 9.8 ± 4.7 | 8.4 ± 3.9 | 2.6 ± 2.7 | 2.7 ± 2.2 | 3.6 ± 2.4 | 2.1 ± 1.8 | 1.7 ± 1.6 |
| Mauls | Starter | 1.2 ± 1.9 | 1.9 ± 1.7 | 1.8 ± 2.1 | 1.6 ± 1.8 | 1.7 ± 1.8 | 0.15 ± 0.43 | 0.17 ± 0.47 | 0.29 ± 0.60 | 0.24 ± 0.49 | 0.22 ± 0.71 |
| | Sub | 0.11 ± 0.32 | 0.37 ± 0.77 | 0.4 ± 0.83 | 0.39 ± 0.61 | 0.23 ± 0.68 | 0.03 ± 0.1 | 0.21 ± 0.58 | 0.16 ± 0.52 | 0.042 ± 0.20 | 0.0 ± 0.0 |
| | Overall | 1.4 ± 1.5 | 2.0 ± 2.04 | 1.9 ± 1.9 | 1.8 ± 0.96 | 1.8 ± 1.4 | 0.15 ± 1.0 | 0.22 ± 0.79 | 0.32 ± 0.83 | 0.24 ± 0.96 | 0.26 ± 0.79 |
| Successful tackles | Starter | 7.2 ± 4.1 | 7.1 ± 4.2 | 9.3 ± 4.6 | 13 ± 5.7 | 11 ± 4.7 | 6.6 ± 3.7 | 8.8 ± 4.3 | 9.4 ± 4.8 | 5.3 ± 2.9 | 3.8 ± 2.5 |
| | Sub | 2.8 ± 2.4 | 3 ± 2.5 | 3.3 ± 2.6 | 4 ± 3 | 4.5 ± 2.9 | 2.2 ± 2 | 3.2 ± 2.5 | 3.2 ± 2.5 | 1.5 ± 1.4 | 1.5 ± 1.5 |
| | Overall | 7.9 ± 3.6 | 9.7 ± 3.8 | 11 ± 3.8 | 14 ± 4.1 | 12 ± 4 | 8.2 ± 3.3 | 9.7 ± 3.5 | 10 ± 4 | 5.5 ± 2.7 | 4.1 ± 2.3 |
| Number of times tackled | Starter | 2.9 ± 2.5 | 4.4 ± 3.2 | 4 ± 2.9 | 5 ± 3.4 | 8.5 ± 4.1 | 3.2 ± 2.5 | 3.3 ± 2.5 | 5.6 ± 3.3 | 5.1 ± 3.1 | 6 ± 3.7 |
| | Sub | 1.2 ± 1.5 | 1.9 ± 2 | 1.4 ± 1.6 | 1.9 ± 1.9 | 3.2 ± 2.4 | 1.3 ± 1.5 | 1.1 ± 1.3 | 1.9 ± 1.8 | 1.7 ± 1.7 | 2 ± 2.1 |
| | Overall | 3.6 ± 2.6 | 6.2 ± 3.2 | 4.7 ± 2.8 | 6.1 ± 3.4 | 9.7 ± 3.9 | 4.3 ± 2.7 | 3.9 ± 2.6 | 6.5 ± 3.1 | 5.4 ± 2.9 | 6.1 ± 3.1 |

*2.5. HAE exposure*

The mean impacts per player match reported in wearable head sensor studies reporting specifically on mean HAE exposure in male American football or rugby union matches has been previously reported.[1] Again, the impact frequency in matches appears to be within a similar region but given the limited rugby union dataset and methodological limitations, a robust comparison between sports was not achievable.[1] More recently, a league and season-wide iMG study on HAE incidence has been conducted in Premiership rugby union.[43] iMG are considered a suitable method to measure on-field HAEs due to their superior coupling to the skull through the upper dentition.[38,44] A similar study in the NFL and National Collegiate Athletic Association (NCAA) Division I subdivisions was undertaken, however HAE incidence was not reported.[45] A smaller-scale study has been published on HAE incidence in an NCAA Division I team using the same iMG system as the Premiership rugby union study.[46] The same iMG system ensures identical signal processing approaches for head kinematic comparisons in the absence of an agreed protocol.[47]

In elite-level American football, the incidence of HAEs above 10 g was 11.2 and 11.3 HAEs per player match for defense and offense, respectively.[46] Incidence of HAEs above 30 g was 1.6 and 2.6 per player match and 0.9 and 1.4 for HAEs above 2.0 krad/s$^2$ for defense and offense, respectively.[46] In elite-level rugby union, the incidence of HAEs above 10 g was 24.0 and 11.9 HAEs per player hour for forwards (defensive) and backs (offensive) players, respectively.[43] Incidence of HAEs above 30 g was 3.0 and 1.3 per player match and 7.3 and 3.6 for HAEs above 2.0 krad/s$^2$ for forwards and backs, respectively.[43] The incidence measures reported by Allan et al.[43] align with those of Tooby et al.[48] on a similar multi-team elite-level rugby cohort. Based on the findings of Quarrie et al.[27] that elite-level rugby union players play on average two-thirds of each match (0.89 hours) in which they appear, HAEs per player match appear higher in rugby union than American football for forward/defensive players based on the peak linear and angular acceleration thresholds above. For back/offensive players, HAEs per player match comparisons between rugby union than American football are less clear based on the peak linear and angular acceleration thresholds above. A percentage of rugby players complete the full duration of a match due to allowable substitutions. The findings from the larger scale NFL and NCAA implementation could provide a more robust comparison. However, the use of a different iMG system, along with its distinct signal processing approach, potentially limits the validity of comparisons.[47,49]

## 3. Discussion

The comparative data presented here indicate notable differences in concussion, HAE exposure, and match scheduling between elite-level rugby union and American football. However, these findings should be interpreted as indicative trends rather than definitive evidence of greater risk in one sport over another. The available datasets differ in their collection methods (e.g., injury reports and secondary data sources), exposure definitions, and reporting frameworks, which may not provide the level of granularity or consistency required for definitive cross-sport comparisons. Robust, independently collected datasets with standardised definitions and methodologies are needed to validate and extend these indicative trends.

Despite these methodological caveats, several patterns emerge that warrant consideration. Rugby union players generally participate in more competitive matches per season and have longer professional careers, which may lead to a greater cumulative exposure to concussion and HAEs over time. Injury surveillance data suggest that rugby union reports a higher number of concussions per match than the National Football League (NFL), and a greater proportion of these occur during training. These findings may reflect differences in match structure, substitution rules, tackling dynamics, or even reporting cultures rather than intrinsic biomechanical risk.

From a biomechanical standpoint, data suggest broadly similar head kinematic profiles for concussion across both sports, but rugby forwards appear to experience higher rates of HAEs above low- and high-magnitude thresholds than comparable defensive positions in American football. This may reflect the continuous nature of rugby play with frequent engagement in tackles, rucks and mauls. However, given the restricted size of the American football iMG utilised, these differences should be viewed as preliminary. A direct comparison of HAE incidence between Allan et al.[43] in elite rugby union and Gabler et al.[45] in elite American football, using harmonised signal-processing protocols, would provide valuable clarity. The integration of iMG technology in both sports represents a critical opportunity to better quantify on-field HAE exposure. Yet, without harmonised analytical frameworks and transparent access to raw data, comparisons across sports, and even within sports, will remain limited. Independent governance of data processing and open-source signal-processing standards would help ensure that future biomechanical insights are scientifically rigorous and comparable.

Beyond biomechanics, the institutional context of concussion and HAE research deserves attention. Many large-scale injury surveillance or iMG initiatives are directed or funded by governing bodies, raising legitimate questions about independence, data ownership, and publication rights. A more transparent, arm’s-length model where funding is administered through independent research councils could enhance credibility and public trust while preserving scientific autonomy.

## 4. Conclusion

Elite-level rugby union and American football are both high-impact contact sports that continue to attract medical and legal scrutiny regarding player safety. This article examined current indicative trends in concussion and HAE incidence across the two sports. Compared with their American football counterparts, rugby union players typically have longer professional careers and participate in more matches each season. Data from the National Football League (NFL) and Rugby Football Union (RFU) injury reports suggest that rugby union records a higher rate of concussions per match and a greater proportion of concussions sustained during training. Emerging biomechanical evidence further indicates that rugby union forwards may experience a higher frequency of HAEs per player match across both lower and higher acceleration thresholds than defensive players in American football. Overall, these patterns point to greater reported exposure among elite rugby union players across several key measures. Collectively, the evidence underscores the importance of independent, transparent, and methodologically rigorous research to advance understanding of concussion and HAE biomechanics and to inform effective player-safety strategies in contact sport.

Data availability: No data are associated with this article

Competing interests: GT has received research funding from Prevent Biometrics and World Rugby. GT previously conducted consultancy work for World Rugby. GT has conducted research with the NFL Player Health and Safety Department via Biocore LLC.

Grant Information: Not applicable

Acknowledgements: Not applicable

Ethics approval and consent to participate: Not applicable

Authors' contributions: Not applicable